\documentclass[prd,preprint,nofootinbib]{revtex4}

\usepackage{graphicx}
\usepackage{amsfonts}
\usepackage{latexsym}
\usepackage{amssymb}
\usepackage{epsfig}

\begin{document}

\title{ Large invisible decay of a Higgs boson to neutrinos }

\author{Osamu Seto}
 \email{seto@physics.umn.edu}
 \affiliation{
 Department of Life Science and Technology,
 Hokkai-Gakuen University,
 Sapporo 062-8605, Japan
}

%

\begin{abstract}
We show that the standard model (SM)-like Higgs boson may decay into neutrinos with
 a sizable decay branching ratio in one well-known two Higgs doublet model,
 so-called neutrinophilic Higgs model. 
This could happen if the mass of the lighter extra neutral Higgs boson is smaller
 than one half of the SM-like Higgs boson mass.
The definite prediction of this scenario is that the rate of the SM-like Higgs boson decay
 into diphoton normalized by the SM value is about $0.9$.
In the case that a neutrino is Majorana particle, 
 a displaced vertex of right-handed neutrino decay would be additionally observed.
This example indicates that a large invisible Higgs boson decay 
 could be irrelevant to dark matter.
\end{abstract}


\preprint{HGU-CAP-038} 

\vspace*{3cm}
\maketitle


\section{Introduction}

The newly discovered particle at the Large Hadron Collider (LHC)
 is now identified as a Higgs boson~\cite{Aad:2012tfa,Chatrchyan:2012ufa}.
Its measured properties such as spin, parity, and couplings
 are consistent with the Higgs boson $h$ in the standard model (SM) of particle
 physics~\cite{Aad:2013wqa,Aad:2013xqa,Chatrchyan:2013iaa,Chatrchyan:2013mxa}
 within uncertainties, which are not very small yet. 
Possible deviations from the SM prediction on the Higgs boson also have
 been examined.
 
One of those is an invisible Higgs boson decay.
Actually, an invisible Higgs boson decay occurs 
 even in the SM through an off-shell $Z$ boson $Z$ pair into four neutrinos $\nu$, as
  $h \rightarrow Z^* Z^* \rightarrow 2 \nu 2 \bar{\nu}$.
Its branching ratio in the SM is of the order of $10^{-3}$.
If once it is found with a larger branching ratio than
 that due to SM processes, this must be a sign of a beyond the SM (BSM).
Such BSM models include, for instance,
 a light neutralino in supersymmetric models~\cite{Griest:1987qv},
 a Majoron~\cite{Joshipura:1992ua}, graviscalars~\cite{Giudice:2000av}, 
 fourth generation neutrino~\cite{Belotsky:2002ym}
 and Higgs portal dark matter~\cite{InvisibleDecayByDM}.
Searches of invisible decays of the Higgs boson $h$ have been carried out
 and to date only an upper bound on the branching ratio of the invisible decay
 has been obtained~\cite{Aad:2014iia,Chatrchyan:2014tja}.

In this paper, we show that the Higgs boson $h$ would decay into four neutrinos
 through an extra Higgs boson,
 which can be seen as the invisible decay, in a class of the two Higgs doublet model (THDM).
The remarkable feature in this scenario is that the invisible final states are
 a SM particle, neutrinos,
 compared with other BSM models mentioned above where final invisible states
 are new hypothetical particles, such as a supersymmetric particle or dark matter.
We consider the so-called neutrinophilic THDM~\cite{Ma,Wang,Nandi},
 where one Higgs doublet provides the mass of the SM fermions,
 while the other generates neutrino Dirac masses
 with its small vacuum expectation value (VEV). 
Phenomenology of the charged Higgs boson was studied
 in Refs.~\cite{Davidson:2009ha,Haba:2011nb}.
In this paper, we will study a possible phenomenology of neutral Higgs bosons
 in those models.
Because of these Yukawa couplings,
 both extra $CP$-even and extra $CP$-odd neutral Higgs bosons, $H$ and $A$, respectively,
 couple mostly with neutrinos.
Thus, through interactions between the SM-like Higgs boson $h$ and
 the extra Higgs bosons $H (A)$, as the $Z$ boson makes new decay processes
\begin{equation}  
h \rightarrow H^{(*)}H^{(*)} \,{\rm or}\, A^{(*)}A^{(*)} \rightarrow 2 \nu 2 \bar{\nu}
\end{equation}
 arise.~\footnote{
This possibility was briefly mentioned in Ref.~\cite{Davidson:2009ha}
 for a very heavy SM-like Higgs boson in a different type of neutrinophilic Higgs model.}
If the intermediate $H$ or $A$ is off shell,
 the resultant contribution is comparable to the SM contribution by the $Z$ boson
 and is not so large.
However, if  either $H$ or $A$ is on shell, the resultant invisible decay width is large.

\section{Neutrinophilic two Higgs doublet model}

The Higgs sector is of the so-called neutrinophilic THDM,
 where one Higgs doublet $\Phi_1$ with
 its VEV $v_1$ generates the mass of the SM fermions,
 while the other $\Phi_2$ generates neutrino Dirac masses through its VEV $v_2 \ll v_1$.
Such a Yukawa coupling is realized by introducing
 the softly broken $Z_2$-parity charge assigned 
 as in Table~\ref{table:parity}. 
\begin{table}[h]
\caption{The assignment of $Z_2$ parity and lepton number.}
\label{table:parity}
\begin{center}
\begin{tabular}{|l|c|c|} \hline
Fields  &  $Z_{2}$ parity & Lepton number \\ \hline\hline
First Higgs doublet, $\Phi_1$  &  $+$ &  0 \\ \hline
Second Higgs doublet, $\Phi_2$ &  $-$ & 0 \\ \hline
Lepton doublet, $L$  &  $+$ &  1 \\ \hline
Right-handed neutrino, $\nu_R$  &  $-$ & $1$ \\ \hline
Right-handed charged lepton, $\ell_R$  &  $+$ & $1$ \\ \hline
Others  &  $+$ & 0 \\ \hline
\end{tabular}
\end{center}
\end{table}
%
%
The Yukawa interaction is given by 
\begin{eqnarray}
 {\cal L}_Y 
   = - y_{\ell_{\alpha}}  \overline{L}_{\alpha} \Phi_1 \ell_{R_{\alpha}}
     - y_{u_{\alpha}}  \overline{Q}_{\alpha} \tilde{\Phi}_1 u_{R_{\alpha}}
     - y_{d_{\alpha}}  \overline{Q}_{\alpha} \Phi_1 d_{R_{\alpha}}
     - y_{\alpha i} \overline{L_{\alpha}} \tilde{\Phi}_2 \nu_{R_i} + {\rm H.c.} , 
 \label{eq:yukawa1}
\end{eqnarray}
where $\tilde{\Phi} = i \sigma_2 \Phi^*$, 
 $Q$ ($L$) is the left-handed $SU(2)$ doublet quark (lepton), and
 $u_R$, $d_R$, $e_R$, and $\nu_R$ are the right-handed (RH) $SU(2)$ singlet fermions, 
 respectively.
$\alpha$ denotes flavor where we neglect mixing in quarks and
 $i$ represents the generation index of RH neutrinos.
If we admit lepton number violation in theory,
 the lepton number violating Majorana mass term
\begin{eqnarray}
 {\cal L}_M  = -\frac{1}{2}\overline{\nu_{R_i}^c} M_i \nu_{R_i}  
 \label{eq:Mmass}
\end{eqnarray}
 also can be introduced~\cite{Ma}.
The scalar potential is given by
\begin{eqnarray}
 V &=& \mu_1^2 |\Phi_1|^2 +\mu_2^2 |\Phi_2|^2 - (\mu_{12}^2
  \Phi_1^\dagger \Phi_2 + {\rm H.c.}) \nonumber \\
 &&+ \lambda_1|\Phi_1|^4   +
  \lambda_2|\Phi_2|^4 + \lambda_3|\Phi_1|^2|\Phi_2|^2 +\lambda_4 |\Phi_1^\dagger \Phi_2|^2 \  +
   \left\{ \frac{\lambda_5}{2} (\Phi_1^\dagger
    \Phi_2)^2 + {\rm H.c.} \right\},
\label{def:potential}
\end{eqnarray}
 where $\mu_{12}$ is the soft breaking parameter of the $Z_2$ parity, as introduced above.
Conditions that the potential(\ref{def:potential}) is bounded from below and a stable vacuum
 are given by~\cite{Kanemura:2000bq}
\begin{eqnarray}
  \lambda_1 > 0, \qquad \lambda_2 > 0, \qquad 2\sqrt{\lambda_1\lambda_2}+\lambda_3 +\min[0, \lambda_4 -|\lambda_5|] > 0.
\end{eqnarray}
Components in two Higgs doublets, each with a VEV, are parameterized as
\begin{eqnarray}
  \Phi_1
   = 
  \left( \begin{array}{c}
          \phi_1^+ \\
          \frac{v_1 + h_1 + i a_1}{\sqrt{2}} \\
         \end{array}
  \right), \hspace{1cm}
%
  \Phi_2
   = 
  \left( \begin{array}{c}
          \phi_2^+ \\
          \frac{v_2 + h_2+ i a_2}{\sqrt{2}} \\
         \end{array}
  \right).
\end{eqnarray}
Following the concept of neutrinophilic Higgs model,
 we take $v_1\simeq v \simeq 246$ GeV and $v_2 \ll v_1$. 
The smallness of $v_2$ is due to the small $\mu_{12}^2$~\cite{Ma}.
We define $\tan\beta = v_1/v_2$ as usual; this corresponds
 to $\tan\beta\gg 1$.
The states $h_1$ and $h_2$ are diagonalized to the mass eigenstates ($h$ and $H$) as 
 \begin{eqnarray}
  \left( \begin{array}{c}
          h_1  \\
          h_2  \\
         \end{array}\right)
   = 
  \left( \begin{array}{cc}
          \cos\alpha & - \sin\alpha\\
          \sin\alpha & \cos\alpha\\
         \end{array}\right)
  \left( \begin{array}{c}
          h \\
          H \\
         \end{array}\right) .
 \end{eqnarray}
Because of $\tan\beta\gg 1$, $\phi_1^+$ and $a_1$ are mostly eaten by the $W$ and $Z$ bosons,
 while we can identify the physical states
 as $H^+ \simeq \phi_2^+$, $A \simeq a_2$, $H \simeq h_2$, and $h \simeq h_1$.
Then, the mixing angle is found to be 
\begin{equation}
 \sin\alpha \simeq \frac{v_2}{v_1} .
\label{Mixing:Higgs}
\end{equation}
Automatically almost, the so-called SM limit $ \sin(\beta-\alpha) = 1 $
 is realized. Because of $\tan\beta \gg 1$, $\alpha \simeq 0$ is realized.
The Higgs boson $h$ with the mass of $125$ GeV is also composed as $ h \simeq h_1$.
From Eq.~(\ref{eq:yukawa1}), the Yukawa interactions of extra neutral Higgs bosons
 are written as
\begin{eqnarray}
{\cal L}_Y &\supset & 
 -\sum_{f=u_i, d_i,\ell_i} \frac{m_f}{v}\frac{\sin\alpha}{\sin\beta}\overline{f} H f -i \frac{m_{u_i}}{v}(-\cot\beta) \overline{u}_i A \gamma_5 u_i
 -i \sum_{f=d_i,\ell_i}\frac{m_f}{v}\cot\beta \overline{f} A \gamma_5 f \nonumber \\
& &-\frac{y_{\alpha i}}{\sqrt{2}} \cos\alpha \overline{\nu_{\alpha}} H P_R \nu_i
   +i \frac{y_{\alpha i}}{\sqrt{2}} \sin\beta \overline{\nu_{\alpha}} A P_R \nu_i
 + \rm{H.c.}
\label{Yukawa}
\end{eqnarray}
We find that $H$ or $A$ decays into mostly neutrinos for
\begin{equation} 
 y_{\alpha i} \gg \frac{\sqrt{2} m_f}{v \tan\beta}.
\label{Cond:Invisible}
\end{equation}
Masses of extra Higgs bosons are given by
\begin{eqnarray}
m_H^2 &=& \mu_2^2+\frac{\lambda_3+\lambda_4+\lambda_5}{2}v^2 , \label{Mass:mH} \\
m_A^2 &=& \mu_2^2+\frac{\lambda_3+\lambda_4-\lambda_5}{2}v^2 , \label{Mass:mA} \\
m_{H^\pm}^2 &=& \mu_2^2+\frac{\lambda_3}{2}v^2 \label{Mass:mcH}  .
\end{eqnarray}
To be consistent with the electroweak precision test,  
 one neutral Higgs boson mass should be close to the charged Higgs boson mass as
\begin{equation}
 m_{H^+} \simeq m_A  \qquad {\rm or} \qquad m_{H^+} \simeq m_H.
\end{equation}
Interactions of extra Higgs bosons with $h$ is 
\begin{eqnarray}
{\cal L} \supset - v \left(  \frac{1}{2}( \lambda_3+ \lambda_4- \lambda_5) A^2
  + \frac{1}{2} ( \lambda_3+ \lambda_4+ \lambda_5 ) H^2+ \lambda_3 |H^+|^2 \right)  h .
\label{HiggsInt}
 \end{eqnarray}
%

\section{Exotic SM-like Higgs boson decay}

Now we consider a case where either $H$ or $A$ is light enough
 to be produced on shell by the $h$ decay.
There are two mass spectra of Higgs bosons that are as consistent
 with the electroweak precision test:
\begin{equation}
m_H  < m_h/2 \ll m_{H^+} \simeq m_A  
\label{spec:ch-A} 
\end{equation}
and 
%
\begin{equation}
m_A  < m_h/2 \ll m_{H^+} \simeq m_H.
\label{spec:ch-H} 
\end{equation}
From the mass formulas(\ref{Mass:mH}), (\ref{Mass:mA}), and (\ref{Mass:mcH}),
 we find that mass spectra (\ref{spec:ch-A}) and (\ref{spec:ch-H}) can be
 realized for 
\begin{eqnarray}
0>\lambda_4 \simeq \lambda_5 
 \quad {\rm and} \quad
 0>\lambda_4 \simeq -\lambda_5 , 
\label{Rel:l4l5}
\end{eqnarray}
 respectively.

\begin{figure}[h,t]
\begin{center}
\epsfig{figure=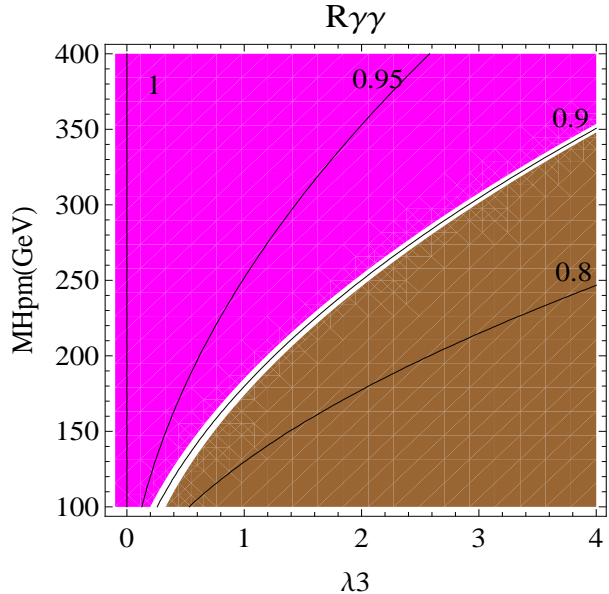, width=8cm,height=8cm,angle=0}
\end{center}
\caption{
Contours of $h \rightarrow \gamma\gamma$ rate
 normalized by the SM value on the $\lambda_3 - m_{H^\pm}$ plane.
In the magenta shaded region, the lighter extra Higgs boson is too heavy to be produced
 by the $h$ decay. 
In the brown shaded region, the extra light Higgs boson is tachyonic.
Here, the extra Higgs boson mass is evaluated with $\lambda_4 =-\lambda_5$. 
 }
\label{Fig:Diphoton}
\end{figure}

With couplings (\ref{Yukawa}) and (\ref{HiggsInt}),
 if $h$ decays into $H$ or $A$, which decays into neutrinos,
 then this fraction is measured as its invisible decay.
\begin{figure}[h,t]
\begin{center}
\epsfig{figure=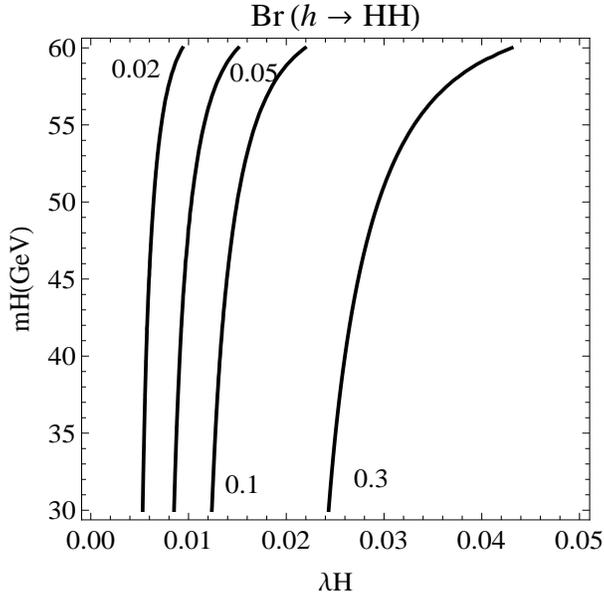, width=8cm,height=8cm,angle=0}
\end{center}
\caption{
Contours of the decay branching ratio of $h$ into $HH$
 in the $\lambda_H - m_H$ plane.
 }
\label{Fig:Dirac:invisible}
\end{figure}
%
The decay width of $h\rightarrow HH$ is given by  
\begin{eqnarray}
\Gamma(h\rightarrow HH ) =
 \lambda_H^2
 \frac{ v^2}{32 \pi m_h } \left( 1 - \frac{4 m_H^2}{m_h^2}\right)^{1/2} ,
\end{eqnarray}
 with $\lambda_H = \lambda_3+ \lambda_4+ \lambda_5$. 
For the case of $h\rightarrow AA$, we obtain the same result
 just by replacing $\lambda_H$ and $m_H$ with $\lambda_A = \lambda_3+ \lambda_4- \lambda_5$ and $m_A$, respectively.
The LHC constraints on exotic decay modes, $h\rightarrow HH$ or $h\rightarrow AA$,
 indicate that $\lambda_H (\lambda_A) \lesssim {\cal O}(10^{-2}) $ is allowed.
We define
\begin{eqnarray}
Br(h\rightarrow HH ) = \frac{\Gamma(h\rightarrow HH )}{\Gamma(h\rightarrow all)},
\end{eqnarray}
 which is shown in Fig.~\ref{Fig:Dirac:invisible}.
By combining the constraint on $\lambda_H (\lambda_A)$ and Eq.~(\ref{Rel:l4l5}),
 we find that, for the light $H$ ($A$),
\begin{equation}
 \lambda_3 \simeq -\lambda_4 -(+) \lambda_5 \label{l3}
\end{equation}
 should be positive and of ${\cal O}(1)$, which leads to the deviation in 
 the diphoton decay rate of $h$ from the SM value~\cite{HHG}, as shown in Fig.~\ref{Fig:Diphoton}.
Here, the brown and magenta shaded regions correspond to the region where
 the extra light Higgs boson is tachyonic and its on shell production is kinematically forbidden, respectively.
One can find that this scenario predicts $R_{\gamma\gamma} \simeq 0.9$.
The signal strength of $h \rightarrow \gamma\gamma$ has been reported as 
 $ 1.17\pm 0.27$ by ATLAS~\cite{Aad:2014eha} and
 $ 1.14{}^ {+0.26}_{-0.23}$ by CMS~\cite{Khachatryan:2014ira}.
%
%

In the following sections, we discuss more  
 detailed phenomenology which depends on neutrino mass nature.

\section{Dirac neutrino case}

The condition (\ref{Cond:Invisible}) is satisfied by a large $\tan\beta$
 for the Dirac neutrino case.
Thus, the light $H$ is, in practice, invisible and 
 we have $Br(h \rightarrow inv)= Br(h\rightarrow HH)$ (shown in Fig.~\ref{Fig:Dirac:invisible}).
The same is true for a light $A$ as well.

The constraint on the charged Higgs boson,
 which decays into a lepton and a neutrino, is, in fact, stringent.
This decay mode is similar to that of a slepton in supersymmetric models~\cite{Davidson:2009ha} and
 masses of the first and the second generation slepton is constrained as
 $m_{\tilde{l}} \gtrsim 300 $ GeV by ATLAS~\cite{Aad:2014vma}
 or $m_{\tilde{l}} \gtrsim 260$ GeV by CMS~\cite{Khachatryan:2014qwa}. 
Although some differences due to its decay branching ratio exist~\cite{Davidson:2009ha},
 roughly speaking, there is a similar bound on $H^+$.
Referring to Fig.~\ref{Fig:Diphoton},
 we find that a rather large coupling $\lambda_3 \gtrsim 3 (2)$ for the ATLAS (CMS) bound
 is required.

\section{Majorana neutrino case}

With the presence of the Majorana mass term (\ref{eq:Mmass}), the neutrino mass matrix is given as
\begin{equation}
{\cal M} = \left(
\begin{array}{cc}
 m_{\nu}^{(loop)} &  \frac{y}{\sqrt{2}} v_2\\
 \frac{y}{\sqrt{2}} v_2  &  M_k 
\end{array}
\right) ,
\end{equation}
 with the radiative generated mass $m_{\nu}^{(loop)}$~\cite{Ma:2006km},
\begin{equation}
m_{\nu}^{(loop)}
 = \sum_k \frac{ y_{\alpha k} y^T_{k\beta} M_k }{16\pi^2}
 \left(
 \frac{m_H^2}{m_H^2-M_k^2} \ln \frac{m_H^2}{M_k^2}  -
 \frac{m_A^2}{m_A^2-M_k^2} \ln \frac{m_A^2}{M_k^2} \right)  .
\label{Mass:neutrino:loop}
\end{equation}
We obtain a light neutrino mass
\begin{equation}
(m_\nu)_{\alpha\beta} = m_{\nu}^{(loop)} - \sum_k \frac{y_{\alpha k} y^T_{k\beta} v_2^2 /2 }{M_k} ,
\label{Mass:neutrino}
\end{equation}
 the mass of a heavier RH-like neutrino $m_{N_R} \simeq M_k$
 and the left-right mixing angle $\theta$~\cite{Type1seesaw}
\begin{equation}
\sin\theta \simeq \frac{y v_2 }{\sqrt{2} M_k} .
\label{Mixing:neutrino}
\end{equation}
$m_{\nu}^{(loop)}$ could be indeed
 dominant ----- or at least comparable with tree level seesaw mass ----- 
 because of $\lambda_5 = \pm {\cal O}(1)$ from Eqs.~(\ref{Rel:l4l5}) and (\ref{l3}).

The charged Higgs boson decays into $cb$ or $tb$,
 depending on the mass in the neutrinophilic Higgs model~\cite{Haba:2011nb}.
The results for the $H^\pm \rightarrow tb$ mode with a normalizing production cross section
 of $1$ pb can be found in Ref.~\cite{CMS:2014pea}. 
However, this constraint is not so stringent
 because the actual production cross section is not so large.
The LHC data constrains the mass of $H^\pm$ decaying
 into $bc$ between $90$ and $150$ GeV~\cite{Aad:2013hla}
 and $H^\pm$ decaying into $\tau\nu$~\cite{Aad:2014kga}.

We note here one cosmological argument on the Majorana neutrino case. 
The lepton number violation by the Majorana nature of neutrino
 plays an important role in cosmology.
Several cosmological discussions on neutrinophilic Higgs model were held
 in Refs.~\cite{HabaSeto,Haba:2013pca,Choi:2012ap}. 
One of them is an enhancement $\Delta L=2$ washout process by large Yukawa couplings
 and relatively light RH neutrinos in a neutrinophilic Higgs model~\cite{HabaSeto}.
Although a discussion of baryogenesis is beyond the scope and purpose of this paper,
 as a necessary condition to have nonvanishing baryon asymmetry in our Universe,
 we roughly evaluate the condition of no strong washout of lepton
 asymmetry,~\footnote{
This is because lepton asymmetry is a potential source of the baryon asymmetry
 in our Universe in the large class of baryogenesis scenario~\cite{FukugitaYanagida}.}
 provided a nonvanishing lepton asymmetry has been generated by any means of
 a higher energy physics process. 
If this condition were violated,
 it would be difficult to explain nonvanishing baryon
 asymmetry in our Universe, because any generated lepton asymmetry is washed out. 
The washout rate is given by $\Gamma_{\Delta L =2} \simeq y^4 T$.
The condition $\Gamma_{\Delta L =2} < H(T) $ at $T \simeq 100$ GeV,
 with $H$ being the cosmic expansion rate, is rewritten as
\begin{equation}
 y \lesssim 10^{-4},
\end{equation}
 which would be regarded as a cosmologically favored region.~\footnote{
One known mechanism of baryogenesis which works without any lepton asymmetry
 is ``baryogenesis via neutrino oscillation''~\cite{Akhmedov:1998qx,Asaka:2005pn}.}

\begin{figure}[h,t]
  \begin{center}
  \begin{tabular}{ccc}
   \includegraphics[width=0.4\textwidth]{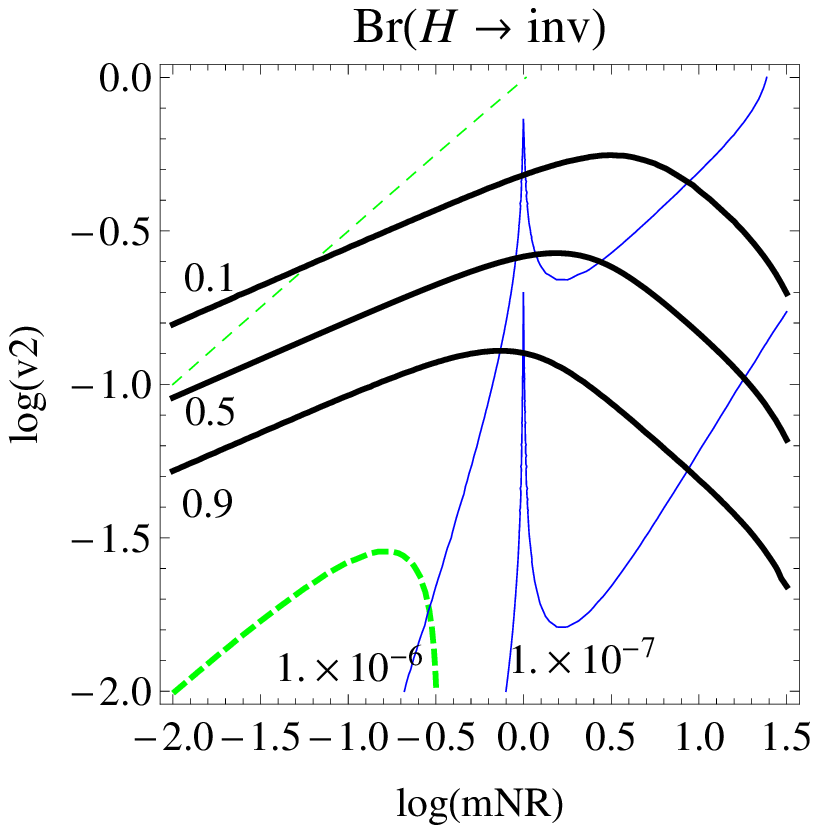}
&\qquad \qquad\qquad &
   \includegraphics[width=0.4\textwidth]{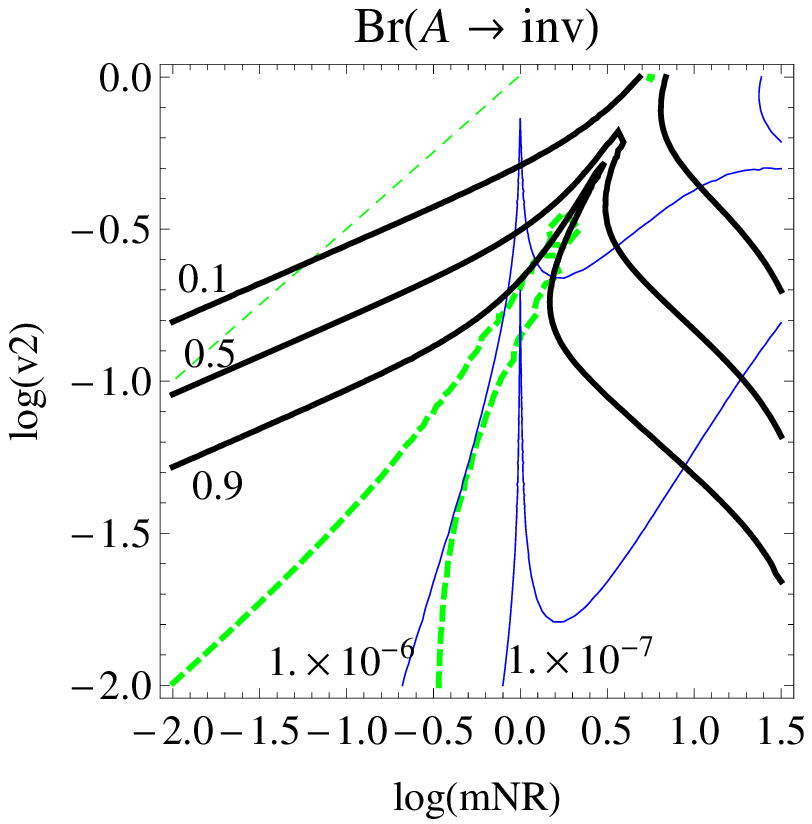}
     \end{tabular}
  \end{center}
\caption{
Contours of the invisible decay branching ratio of the lighter extra neutral Higgs boson,
 $H$ or $A$, for $m_{\nu} = 0.05$ eV on the $\log_{10}(m_{N_R}/{\rm GeV}) - \log_{10}(v_2/{\rm GeV})$ plane.
The thin blue lines are contours of $\sin\theta$.
For the dashed lines, see the text. 
Here we take $m_H, m_A = 60$ and $200$ GeV (left panel) and $200$ and $60$ GeV (right panel).
 }
\label{Fig:nuYukawa}
\end{figure}

Now we discuss the decay of $H$ or $A$.
For $m_{N_R} < m_{H/A}$, an extra neutral Higgs boson $H (A)$ decay produces
 one light left-handed-like neutrino $\nu$
 and the other heavy RH-like neutrino $N_R$.
The amplitude is calculated as
\begin{eqnarray}
\overline{ |{\cal M}(H/A \rightarrow \nu N_R)|^2 } = 2 |y|^2  (p_1\cdot p_2) 
 = |y|^2 ( (p_1+p_2)^2 - m_{N_R}^2) ,
 \end{eqnarray}
 where $p_1$ and $p_2$ are outgoing momentum of $\nu$ and $N_R$, respectively.
In addition, $m_{\nu}$ is neglected and indexes of $y$ are omitted.
The decay width is given by  
\begin{eqnarray}
\Gamma(H/A \rightarrow \nu N_R) &=& \frac{1}{16\pi m_{H/A}{}^3} \sum |y|^2  ( m_{H/A}{}^2 - m_{N_R}^2)^2. 
\end{eqnarray}
Here, the summation $\sum$ is taken for all kinematically possible modes.
An extra Higgs boson decays into SM fermions, mostly the bottom quark, through a tiny mixing $\alpha$.
Thus, its decay width is strongly suppressed by a large $\tan\beta$ as
\begin{eqnarray}
\Gamma(H/A \rightarrow b\bar{b}) \simeq \frac{3}{8\pi} \left(\frac{m_b}{v \tan\beta}\right)^2 m_{H/A}. 
\end{eqnarray}
Here, we define 
\begin{equation}
Br(H/A \rightarrow inv) = \frac{\Gamma(H/A \rightarrow \nu N_R)}
 {\Gamma(H/A \rightarrow b\bar{b}) + \Gamma(H/A \rightarrow \nu N_R) }, 
\end{equation}
and
\begin{equation}
Br(h \rightarrow 2H/2A \rightarrow 2\nu 2 N_R) = Br(h \rightarrow HH/AA) Br(H/A \rightarrow inv) .
\end{equation}
Figure~\ref{Fig:nuYukawa} shows the contour plot of the invisible decay branching ratio
 with thick black lines
 as well as the contour of Eq.~(\ref{Mixing:neutrino}) with thin blue thin lines
 of $H$ and $A$, respectively. 
In both cases, the invisible decay branching ratio is large for $v_2 < 0.1$ GeV.
The dashed green (thick) lines are contours of the typical size of Yukawa coupling
 $y=10^{-5} (10^{-4})$ estimated from Eq.~(\ref{Mass:neutrino}) with
 the atmospheric neutrino mass.
As discussed above, $y \simeq 10^{-4}$ would be critical
 when we consider nonvanishing baryon asymmetry in our Universe.
In both panels, neutrino masses dominantly come from the tree level seesaw
 at the upper left region, and do from $m_{\nu}^{loop}$
 at the lower right region.
In the light $A$ case shown in the right panel of Fig.~\ref{Fig:nuYukawa},
 the destructive cancellation between $m_{\nu}^{loop}$ and the seesaw term takes place.
This cancellation makes cuspy curves of the invisible branching ratio
 and Yukawa couplings in contours.
For the $H$ decay, $m_H=60$ GeV and $ m_A = 200$ GeV are taken.
For the $A$ decay, $m_H=200$ GeV and $m_A = 60$ GeV are taken.
%
%

\begin{figure}[h,t]
\begin{center}
\epsfig{figure=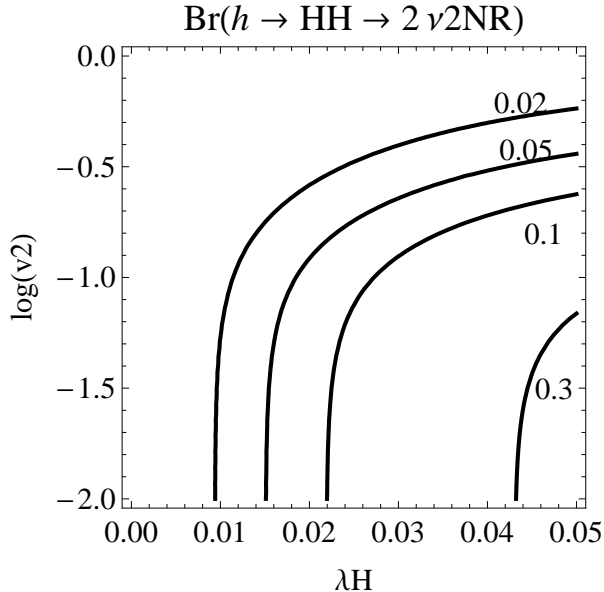, width=8cm,height=8cm,angle=0}
\end{center}
\caption{
Contours of the decay branching ratio of the SM-like Higgs boson, $Br(h \rightarrow 2H \rightarrow 2 \nu 2 N_R)$,  
 in the $\lambda_H - v_2$ plane. Here, we take $m_{N_R} = 10$ GeV.
 }
\label{Fig:Majorana:invisible}
\end{figure}

Figure~\ref{Fig:Majorana:invisible} is the contour plot of
 the invisible decay branching ratio of $h$.
Here, $m_H= 60$ GeV and $m_{N_R} = 10$ GeV are taken.
%
%

The produced RH neutrino $N_R$ decays as $N_R \rightarrow Z^* \nu , h^* \nu$ or $ W^* \ell$ through
 a tiny left-right mixing of $ \sin\theta \lesssim 
 {\cal O}( 10^{-6} )$.
Here, a sign of inequality becomes more appropriate as $m_{\nu}^{(loop)}$ becomes sizable.
For such a left-right mixing of the order of $10^{-6}$ or less, 
 the displaced vertex of $N_R$ decay could be generated, and
 the decay length of the RH neutrino becomes 
 $c \tau_{N_R} \gtrsim 1$ cm for $m_{N_R} = {\cal O}(10)$ GeV~\cite{Displaced}. 
For a further lighter $m_{N_R}$ or a much smaller $\sin\theta$,
 $N_R$ would not decay inside the detector. 
One can see that such a small mixing is realized in Fig.~\ref{Fig:nuYukawa}.

On the other hand, for  $m_{N_R} > m_{H/A}$, $H$ or $A$ decays
 into SM fermions through a tiny mixing of a Higgs boson (\ref{Mixing:Higgs})
 as in the usual type-I THDM. 
%

\section{Summary}

We have shown that
 the SM-like Higgs boson could have a sizable invisible decay
 branching ratio such as ${\cal O}(10) \% $
 with four neutrinos final states, $h \rightarrow 2 \nu2\bar{\nu}$ for a Dirac neutrino
 and $h \rightarrow 2 \nu 2 N_R$ for a Majorana neutrino,
 in neutrinophilic Higgs doublet models,
 if one of the extra Higgs bosons is light enough to be produced
 by the SM-like Higgs boson decay. 
For the Majorana neutrino, this becomes a case in the parameter region $v_2 \lesssim 0.1$ GeV.
Because of this mass spectrum of Higgs bosons,
 the SM normalized decay rate of $h \rightarrow \gamma\gamma$ is predicted to be $0.9$. 
In the Majorana neutrino case, the displaced vertex of a $N_R$ decay also would be observed.
Although the invisible decay of the Higgs boson was recently widely discussed
 in~\cite{InvisibleDecayByDM} or applied to
 dark matter physics~\cite{Aad:2014iia,Chatrchyan:2014tja},
we emphasis that such a size of invisible decay can be realized just within
 simple THDM without a dark matter candidate.


\section*{Acknowledgments}
We would like to thank Koji Tsumura and Shigeki Matsumoto for the valuable comments.
This work is supported in part by Grant-in-Aid for Scientific Research 
 No.~2610551401 from
 the Ministry of Education, Culture, Sports, Science and Technology in Japan.
%


\appendix




\end{document}